\documentstyle[preprint,pre,aps,eqsecnum]{revtex}
\draft
\begin{document}
\draft
\title{VARIATIONAL SCHEMES IN THE \\
FOKKER-PLANCK EQUATION}
\author{T. Blum and A. J. McKane}
\address{Department of Theoretical Physics \\
University of Manchester\\
Manchester M13 9PL, UK}
\date{\today}
\maketitle
\begin{abstract}
We investigate variational methods for finding approximate solutions
to the Fokker-Planck equation, especially in cases lacking detailed
balance.
\ These schemes fall into two classes: those in which a Hermitian
operator is constructed from the (non-Hermitian) Fokker-Planck operator,
and those which are based on soluble approximations to this operator.
\ The various approaches are first tested on a simple quantum-mechanical
problem and then applied to a toy Fokker-Planck equation.
\ The problem of a particle moving in a potential and subject to
external non-white noise is then investigated using the formalism
developed earlier on in the paper.
\end{abstract}
\section{Introduction}
Variational schemes, such as the Rayleigh-Ritz procedure \cite{Schiff}
are well known methods for finding approximate solutions to the
Schr\"odinger equation.
\ There has been far less work done in applying analogous procedures
to the Fokker-Planck equation, even though the equations share a
similar structure.
\ One of the reasons for this situation is that the differential
operator in the Fokker-Planck equation is not, in general, self-adjoint,
making the formalism more complicated: \ eigenvalues are not
necessarily real, nor are the right and left eigenfunctions equal.
\ Another reason is the existence of a zero eigenvalue, corresponding to
the steady-state of the system.
\ Under certain conditions (the ``potential conditions") the determination
of the steady-state probability distribution function reduces to
quadratures.
\ Since all eigenvalues have a non-negative real part \cite{Risken}, in
this case the lowest eigenvalue and eigenfunction are known exactly, and
approximation schemes have only to be developed for ``excited" states.
\ The formalism for this has been developed and applied to a number of
problems \cite{Risken,Brand}.
\ However, for many systems of interest the potential conditions do not
hold, and as a consequence the steady-state distribution cannot be
determined in closed form.
\ It is desirable to have variational schemes in this case as well.
\ Some work has been carried out by Seybold \cite{Sey}, but no systematic
discussion exists in the literature.
\ The purpose of this paper is to investigate the usefulness of a number
of variational schemes in the case where detailed balance does not hold.
\ These are illustrated on simple two-dimensional systems, which are the
simplest systems for which potential conditions need not hold.

The outline of the paper is as follows.
\ In section two we review some general formalism for the
Fokker-Planck equation and in addition discuss the potential
conditions and show how they lead to an exact solution for the
stationary probability distribution.
\ Section three introduces the variational methods that are the heart
of this work
\ In section four, we consider some examples of these schemes in
action, including their application to the colored-noise problem.

\section{Formalism}

In this section we first present some of the formalism associated with
the Fokker-Planck equation and then treat the potential conditions,
demonstrating how they lead to an exact solution for the steady-state
probability distribution.
\ Here, we will see that making the Fokker-Planck equation look as much
as possible like quantum mechanics is very natural.
\subsection{General Formalism}
The Fokker-Planck equation for a system with $N$ degrees of freedom is
\cite{Risken,Gardiner}
\begin{equation}
\partial_t ~p(\underline{x},t) \ = \ L ~p(\underline{x},t),
\end{equation}
where $L$ is a differential operator of the form:
\begin{equation}
L \ = \ - \partial_i A_i(\underline{x})
\ +\  \partial_i \partial_j B_{ij}(\underline{x}),
\label{operator}
\end{equation}
where the summation convention is understood.
\ Here $A_i(\underline{x})$ and $B_{ij}(\underline{x})=B_{ji}(\underline{x})$
are the drift vector and the diffusion matrix respectively, $\partial_i$
means $\partial /\partial x_i$ and $i,j=1,2,...,N$.
\ Looking for separable solutions of the form:
\begin{equation}
p(\underline{x},t) \ = \ P_{n} (\underline{x})~
\exp \{ -\lambda_n t \}
\label{sep}
\end{equation}
leads to the eigenfunction equation:
\begin{equation}
L ~P_{n} (\underline{x}) \ = \ -~ \lambda_n ~
P_{n} (\underline{x}),
\label{eigen}
\end{equation}
where we have assumed for convenience a discrete eigenvalue
spectrum with $n=0,1,2,....$
\ Since $L$ cannot, in general, be brought into Hermitian form, the
eigenfunctions $Q_{m} (\underline{x})$ of the adjoint operator
have to be found:
\begin{equation}
L^{\dag}~ Q_{m} (\underline{x}) \ = \  -~ \lambda_m ~
Q_{m} (\underline{x}),
\label{adjoint}
\end{equation}
as well.
\ The set of eigenvalues in (\ref{eigen}) and (\ref{adjoint}) are
equal; moreover, since the operator is real, the complex conjugate of
an eigenvalue is also an eigenvalue.
\ The set of functions $P_{n}(\underline{x})$ and
$Q_{m}(\underline{x})$ are bi-orthogonal:
\begin{equation}
\int d^{N}\underline{x} ~P_{n}(\underline{x})~
Q_{m}(\underline{x}) \ = \ \delta_{n,m}.
\label{orth}
\end{equation}
Equation (\ref{sep}) shows that the stationary state corresponds to
$\lambda_0 = 0$ i.e. $P_0 (\underline{x}) = p_{st} (\underline{x})$.
\ From the form of $L$ we see that $Q_{0} (\underline{x})$ is a
constant, and then from (\ref{orth}) that $Q_0 (\underline{x}) = 1$.
\ Henceforth, we will frequently denote operations such as those
found in eqs. (\ref{eigen}), (\ref{adjoint}) and (\ref{orth}) in the
following notation:
\begin{eqnarray}
&L~|P_{n} \rangle \ &=\ -\lambda_n ~|P_{n} \rangle
\nonumber \\
&\langle Q_{m} |~L \ &=\ -\lambda_m ~\langle Q_{m} |
\nonumber \\
&\langle Q_{m} | P_{n} \rangle \ &= \ \ \ \delta_{n,m}.
\label{bra-ket}
\end{eqnarray}

\subsection{The Potential Conditions}
The study of the Fokker-Planck equation simplifies considerably if
the drift vector $A_i (\underline{x})$ and the diffusion matrix
$B_{ij} (\underline{x})$ satisfy the so-called potential conditions
mentioned in Section I.
\ There are various ways to introduce these conditions, but perhaps
the simplest is first to note that the Fokker-Planck equation may
be written in the form of a continuity equation: \
$\partial _{t} p(\underline{x},t) + \partial _{i} J_{i}
(\underline{x},t) = 0$, where $J_{i}$ is:
\begin{equation}
J_i(\underline{x})\ =\ \left[A_i(\underline{x}) - \partial_j
B_{ij}(\underline{x}) \right] p(\underline{x},t)
\label{probcurrent}
\end{equation}
and called the probability current.
\ Stationarity implies $\nabla \cdot \underline{J} = 0$, but if it is
also true that $\underline{J} =
\underline{0}$, then
\begin{equation}
\partial_j \left[ B_{ij} (\underline{x})~ p_{\rm st} (\underline{x})
\right] \ = \ A_i (\underline{x}) ~p_{\rm st} (\underline{x}).
\label{current}
\end{equation}
(It turns out that an equivalent statement is
$P_{n}(\underline{x})=p_{\rm st}(\underline{x}) Q_{n}(\underline{x})$
for all $n$.)
\ If in addition the diffusion matrix $B$ has an inverse then
\begin{equation}
C_i (\underline{x}) \ \equiv \ \partial _i \ln p_{\rm st} (\underline{x})
\ = \ B^{-1}_{ij} (\underline{x}) \left[ A_j (\underline{x}) -
\partial _k B_{jk} (\underline{x}) \right]
\label{potential}
\end{equation}
is a gradient.
\ Necessary and sufficient conditions for this are:
\begin{equation}
\frac{\partial C_i}{\partial x_j} \ =\  \frac{\partial C_j}{\partial x_i},
\label{conditions}
\end{equation}
with $i,j = 1,...,N$.
\ Provided these conditions hold, the determination of the stationary
probability distribution is reduced to quadratures:
\begin{equation}
p_{\rm st} (\underline{x}) \ = \ \exp \left\{
\int^{\underline{x}} dx'_i ~ C_i (\underline{x'}) \right\}.
\label{quadratures}
\end{equation}
Furthermore when (\ref{current}) holds, the operator
\begin{equation}
{{\cal L}} = \left(p_{\rm st} (\underline{x}) \right)^{-1/2}~ L~ \left(p_{\rm
st}
(\underline{x}) \right)^{1/2}
\label{rotation}
\end{equation}
is Hermitian \cite{Risken} with the real eigenfunctions
\begin{equation}
\psi _{n} (\underline{x}) \ = \ \frac{P_{n}(\underline{x})}
{\left(p_{\rm st} (\underline{x}) \right)^{1/2}} \ = \
\left(p_{\rm st} (\underline{x}) \right)^{1/2}
{}~Q_{n} (\underline{x}),
\label{pandq}
\end{equation}
and the eigenvalues are non-negative.

Of course, these simplifications occur only in exceptional cases; for
most cases the operator $L$ cannot be brought into Hermitian form as
in (\ref{rotation}).
\ Nevertheless, it is still useful to define the operator ${{\cal L}}$, since
it turns out that splitting this operator (as opposed to $L$) up into
Hermitian and anti-Hermitian parts is natural in the sense that it is
dictated by the underlying temporal symmetries of the system
\cite{Risken}.
\ This decomposition can be achieved explicitly by writing $A_i =
A^{+}_i + A^{-}_i$, where $A^{+}_i$ is defined by $A^{+}_{i}=
p_{\rm st}^{-1} \partial _j[B_{ij}p_{\rm st}]$.
\ Denoting the operator of the form (\ref{operator}), with $A$ replaced
by $A^{+}$, $L$ by $L^{+}$ and defining $L^{-} = -\partial _i A^{-}_i$,
the original Fokker-Planck operator may be written as $L = L^{+} +
L^{-}$ with the corresponding ${{\cal L}}$ operators defined by
(\ref{rotation}) having a similar decomposition: \ ${{\cal L}} =
{{\cal L}}^{+} + {{\cal L}}^{-}$.
\ By construction ${{\cal L}}^{+}$ is Hermitian, since $A^{+}$ satisfies the
conditions (\ref{current}).
\ It is also easy to check that ${{\cal L}}^{-}$ is anti-Hermitian using
$L^{-} p_{\rm st} (\underline{x}) = 0$.

In general, the operators ${{\cal L}}$ and ${{\cal L}}^{\dag}$ will have
a different set of (complex) eigenfunctions: from (\ref{eigen}),
(\ref{adjoint}) and (\ref{rotation}) the eigenfunctions of these
operators are:
\begin{equation}
\psi_{n} (\underline{x}) \ =\  \frac{P_{n}(\underline{x})}
{\left(p_{\rm st} (\underline{x}) \right)^{1/2}} \ \ \ {\rm and}
\ \ \ \chi _{n} (\underline{x})\ = \ \left(p_{\rm st}
(\underline{x}) \right)^{1/2} Q_{n} (\underline{x}).
\label{pandc}
\end{equation}
{}From (\ref{orth}) they satisfy the orthonormality condition
\begin{equation}
\langle \chi_{m}|\psi_{n} \rangle \ = \
\int d^{N}\underline{x} ~\psi _{n}(\underline{x})
{}~\chi _{m}(\underline{x}) \ = \ \delta_{n,m}.
\label{normal}
\end{equation}
The eigenvalues, although complex, have a non-negative real part
\cite{Risken,Graham} as expected on physical grounds.
\ The eigenfunction of both ${{\cal L}}$ and ${{\cal L}}^{\dag}$
corresponding to the zero eigenvalue is
$\left(p_{\rm st} (\underline{x}) \right)^{1/2}$.

\section{Variational Approaches}

Variational approaches are quite useful for finding ground-state
energies in quantum mechanics because: \  i) an error of order $\epsilon$
in the variational wave function results in an error of order
$\epsilon^2$ in the variational energy; and ii) the true ground-state
energy is known to be lower than the variational energy.
Since the lowest eigenvalue of a Fokker-Planck operator is
identically zero, the focus becomes the eigenstates (the lowest being
the stationary probability distribution) and/or excited eigenvalues
(the first excited eigenvalue being the reciprocal of the longest
relaxation time of the system \cite{Brand}).

Some work on Fokker-Planck variational approaches has been carried out,
mainly on one-component problems.
\ In such cases the potential conditions automatically hold, the stationary
probability distribution is known exactly, and the interest
was naturally centered on higher eigenvalues.
\ In these circumstances, progress is made by obtaining the Hermitian
operator ${{\cal L}}$ and proceeding as in quantum mechanics.
\ In calculations of the lowest non-zero eigenvalue, the nice features
of a variational approach can be recovered simply by ensuring that
the variational state is orthogonal to the stationary probability
distribution.
\ However, since most problems do not obey the potential conditions,
variational schemes applicable to these more general situations need
to be developed.

\subsection{Constructing a Hermitian Operator: \ The $L^{\dagger}L$
Approach}
When the potential conditions do not hold, one can still mimic quantum
mechanics by constructing a self-adjoint operator from the Fokker-Planck
operator $L$.
\ A natural choice is $L^{\dagger}L$ since it shares right
eigenstates with $L$ and its eigenvalues $\Lambda_n$ are related
to those of $L$ through $\Lambda_n=\lambda_n^{*}\lambda_n$.
\ With a Hermitian operator in hand one simply proceeds as in
quantum mechanics.
\ In addition to the drawback that there is no longer any interest in
finding the lowest eigenvalue (it is identically zero), another disadvantage
of this scheme compared to
Rayleigh-Ritz in quantum mechanics, is that the calculation requires
twice as many operations --- first operating with $L$ and then with
$L^{\dagger}$, rather than just with the Hamiltonian $H$.

It is natural to consider minimizing
$\langle P_0^{\rm tr}|L^{\dagger} L|P_0^{\rm tr}
\rangle/\langle P_0^{\rm tr}|P_0^{\rm tr} \rangle$ \cite{Sey}, where
$P_0^{\rm tr}(\underline{x})$ is a trial distribution, as a means for
finding a reasonable approximation to the steady-state distribution.
\ This quantity is bounded from below by zero and equals zero only if
the choice $P_0^{\rm tr}(\underline{x})$ is equal to the exact
steady-state distribution $P_0(\underline{x})$.

Note that when examining the eigenvalue equation $L P_n =
-\lambda_n P_n$ one has the freedom to transform the equation to
$\tilde L \tilde P_n = -\lambda_n \tilde P_n$ where $\tilde L =
R^{-1/2}LR^{1/2}$ and $\tilde P_n=R^{-1/2}P_n$ provided $R$ has
no zeros.
\ Hence, one alternative would be to minimize
$\langle \tilde P_0^{\rm tr}|\tilde L^{\dagger} \tilde L|\tilde P_0^{\rm tr}
\rangle/\langle \tilde P_0^{\rm tr}|\tilde P_0^{\rm tr} \rangle$.
\ Calculational convenience sometimes dictates using $R=P_0^{\rm tr}$.

The first excited eigenvalue $\lambda_1$ (the reciprocal of the longest
relaxation time of the system) is accessible to a straightforward
variational approach provided $|P_1\rangle$ has a different symmetry
than $|P_0 \rangle$.
\ In such cases $\langle P_1^{\rm tr}|L^{\dagger}L |P_1^{\rm tr} \rangle
/\langle P_1^{\rm tr} |P_1^{\rm tr} \rangle$ yields an upper bound on
$\Lambda_1$ provided one can guarantee that $|P_1^{\rm tr}\rangle$ is
orthogonal to $|P_0\rangle$.

It is possible to lower the bounds on $\Lambda_n$ by improving
$|P_n^{\rm tr}\rangle$
(making a better choice or one with more variational parameters).
\ A systematic procedure, which is similar to Lan\c czos tridiagonalization
\cite{Dagotto}, uses:
\begin{equation}
|{\cal P}_{\rm tr}\rangle \ = \  |P_{\rm tr}\rangle
+ \alpha L|P_{\rm tr}\rangle +
\beta L^2|P_{\rm tr}\rangle + \ldots,
\end{equation}
as the improved trial distribution, where $\alpha,\beta,...$ are
additional variational parameters.
\ Finally we note that it is possible to get some sense of how
good the variational estimate is, since in the
Fokker-Planck problem the lowest eigenvalue should be zero.
\ In addition, for the higher eigenvalues there exists a
procedure which finds lower as well as upper bounds on
the eigenvalues \cite{Brand,Weinstein}.

\subsection{Variation within Perturbation}
The other variational approaches we will discuss have their foundations
in perturbation theory.
\ If the Fokker-Planck operator at hand is in some sense close to
one which is solvable, then a perturbation approach is a viable
scheme of approximation.
\ One can make a perturbative expansion variational by introducing
parameters into the solvable part and eventually choosing them to
minimize some measure of the ``nearness'' of the two problems.

The perturbative approach begins by splitting the Fokker-Planck
operator as follows:
\begin{equation}
L \ = \ L^{(0)} \  + \ \epsilon~L^{(1)},
\label{perturb}
\end{equation}
where $L^{(0)}$ is the operator of some completely solvable problem,
i.e., all of its eigenvalues $\lambda_n^{(0)}$ as well as its right and
left eigenfunctions, $P_{n}^{(0)}(\underline x)$ and
$Q_{m}^{(0)}(\underline x)$ are known.
\ The variational aspect of this approach requires $L^{(0)}$ to
contain some as yet undetermined parameters.
\ Note that $\epsilon$ need not be small and is used here primarily
as a counting device.
\ Expanding the eigenfunctions and eigenvalues of $L$ as a series
in $\epsilon$ yields:
\begin{equation}
\left( L^{(0)}+ \epsilon L^{(1)}\right)
\left(|P_{n}^{(0)}\rangle + \epsilon |P_{n}^{(1)} \rangle
+\ldots \right)\ =\ -\left(\lambda_n^{(0)}+\epsilon
\lambda_n^{(1)}+\ldots \right) \left(|P_{n}^{(0)}\rangle
+\epsilon |P_{n}^{(1)}\rangle +\ldots \right).
\label{perturb2}
\end{equation}
Manipulating the terms of order $\epsilon$ is the usual ways leads
to:
\begin{eqnarray}
\lambda_n^{(1)}&\ = \ &-\epsilon ~\langle Q_{n}^{(0)}|L^{(1)}|
P_{n}^{(0)} \rangle
\\
|P_{n}^{(1)} \rangle& \ = \ & -\epsilon \sum_{m \neq n}
\frac{| P_{m}^{(0)} \rangle \langle Q_{m}^{(0)} |
L^{(1)} | P_{n}^{(0)} \rangle }{(\lambda_n^{(0)} -
\lambda_m^{(0)})} \\
\langle Q_{n}^{(1)} |& \ = \ &- \epsilon \sum_{m \neq n}
\frac{  \langle Q_{n}^{(0)} | L^{(1)} | P_{m}^{(0)}
\rangle \langle Q_{m}^{(0)}|}{(\lambda_n^{(0)} - \lambda_m^{(0)})}.
\label{first-order2b}
\end{eqnarray}
The terms of order $\epsilon^2$ lead to:
\begin{equation}
\lambda_n^{(2)}\ = \ \epsilon^2\sum_{m\neq n}
{\langle Q_{n}^{(0)}|L^{(1)}|P_{m}^{(0)} \rangle
\langle Q_{m}^{(0)}|L^{(1)}|P_{n}^{(0)} \rangle
\over (\lambda_n^{(0)}-\lambda_m^{(0)}) },
\label{second-order-a}
\end{equation}
and so on.

Notice that the structure of the perturbation expansion is such that
$\lambda_0=0$ and $\langle Q_0|=1$ at every order, as these results are
exact.
\ Furthermore, it is worth remarking that the expansion for $|P_0\rangle$
does not necessarily remain everywhere positive order-by-order; thus
there can arise difficulties in interpreting a truncated expansion of
$|P_0\rangle$ as a probability distribution.
\ Another feature of the perturbation expansion to note is that if the
$\lambda_n^{(0)}$'s are all real, then the perturbation expansion for the
eigenvalues remains real (provided there are no degenerate
$\lambda_n^{(0)}$'s which would necessitate degenerate perturbation
theory); that is, the imaginary part of the eigenvalue is inaccessible
to a perturbation theory that begins with purely real eigenvalues.

If $L_0$ has been chosen so as to satisfy the potential conditions, then it
is convenient to consider the perturbative expansion of the operator
transformed by $P_0^{(0)}$:
\begin{equation}
\left(\ell^{(0)} + \epsilon \ell^{(1)} \right)
\left(|\psi_n^{(0)}\rangle+\epsilon |\psi_n^{(1)}\rangle
 +\ldots\right) \ =\ -\left( \lambda_n^{(0)}+\epsilon
\lambda_n^{(1)} + \ldots \right) \left(| \psi_n^{(0)} \rangle
+ \epsilon | \psi_n^{(1)} \rangle + \ldots \right)
\label{pzero-rot}
\end{equation}
where
\begin{eqnarray}
&\ell^{(0)}\ =\ &\left(P_0^{(0)}(\underline{x})\right)^{-1/2}
L^{(0)} \left(P_0^{(0)}(\underline{x})\right)^{1/2}
\nonumber \\
&\ell^{(1)}\ =\ &\left(P_0^{(0)}(\underline{x})\right)^{-1/2}
L^{(1)} \left(P_0^{(0)}(\underline{x})\right)^{1/2}
\nonumber \\
&\psi_n^{(j)}(\underline{x}) \ = \ &\ \ \ {P_{n}^{(j)}({\underline
x}) \over \left( P_{0}^{(0)}({\underline x}) \right)^{1/2}}.
\end{eqnarray}
This is because the left and right eigenstates at zeroth order are
identical, which is of considerable calculational convenience.

Of course, the more natural rotation is by the true stationary
probability distribution, but that is unknown.
\ In the perturbative approach it is possible to perform this
transformation order-by-order as follows:
\begin{eqnarray}
{{\cal L}} \ = \ &&\left( P_0^{(0)} + \epsilon P_0^{(1)} +
\ldots \right)^{-1/2}
\left( L^{(0)}+\epsilon L^{(1)} \right)
\left( P_0^{(0)}+ \epsilon P_0^{(1)} + \ldots \right)^{1/2}
\nonumber \\
{{\cal L}} \ = \ && {{\cal L}}^{(0)} + \epsilon {{\cal L}}^{(1)}
+ \epsilon^2 {{\cal L}}^{(2)} + \ldots
\label{curly-expansion}
\end{eqnarray}
where
\begin{equation}
{{\cal L}}^{(0)} \ = \ \left( P_0^{(0)} \right)^{-1/2}
L^{(0)} \left( P_0^{(0)} \right)^{1/2}
\label{curly-l-zero}
\end{equation}
\begin{equation}
{{\cal L}}^{(1)} \ = \ \left( P_0^{(0)} \right)^{-1/2} \left(
L^{(1)} + \frac{1}{2} \left[ L^{(0)}, \frac{P_0^{(1)}}{P_0^{(0)}}
\right] \right) \left( P_0^{(0)} \right)^{1/2},
\label{curly-l-one}
\end{equation}
and so on.
\ Then one could proceed with
\begin{equation}
\left( {{\cal L}}^{(0)} + \epsilon {{\cal L}}^{(1)} + \ldots \right)
\left( |\Psi_n^{(0)} \rangle + \epsilon |\Psi_n^{(1)} \rangle
+ \ldots \right)
\ =\
-\left(\lambda_n^{(0)} + \epsilon \lambda_n^{(1)} + \ldots \right)
\left( |\Psi_n^{(0)} \rangle + \epsilon |\Psi_n^{(1)} \rangle
+ \ldots \right).
\end{equation}
This approach is rather cumbersome, especially when it is noted that
finding ${{\cal L}}^{(1)}$ requires knowing $P_0^{(1)}$.

Since the lowest eigenvalue in the Fokker-Planck perturbation expansion
is zero order-by-order, some other quantity must be found to vary.
\ It is desirable, though not necessary, to vary a bounded expression,
as the bound helps to ensure a sensible result.
\ This motivation led us to consider varying $\langle P_0^{(1)}
| P_0^{(1)} \rangle$ or alternatively $\langle \Psi_0^{(1)} |
\Psi_0^{(1)} \rangle$ which is positive semi-definite by construction.
\ We call this approach the ``minimal corrected wave function"
criterion or MCW.
\ The variation seeks to minimize $\langle P_0^{(1)}| P_0^{(1)} \rangle$
--- the philosophy being that the closer $|P_0^{(0)}\rangle$ is to the
actual $p_{st}(\underline x)$, the smaller its correction should be.
\ The procedure might be continued by minimizing $\langle P_0^{(2)}
| P_0^{(2)} \rangle$ to determine the parameters to be inserted
in $|P_0^{(0)}\rangle+|P_0^{(1)}\rangle$, and so on.
\ To calculate $\lambda_1$, $\langle P_1^{(n)} | P_1^{(n)} \rangle$ could
be minimized to fix the parameters to be used in $\lambda_1^{(0)}
+ \lambda_1^{(1)} + ... + \lambda_1^{(n)}$.

Another method based on perturbation theory with undetermined
parameters has been used by Edwards and co-workers on problems such as
polymers with excluded volume \cite{Edw-Singh} and the Fokker-Planck
formulation of the KPZ equation \cite{Schwartz-Edw}.
\ Stevenson has dubbed it the ``fastest apparent convergence" criterion
or FAC \cite{Stevenson}.
\ First a number of terms in a perturbative expansion of the quantity of
interest are calculated.
\ FAC then assumes that the zeroth-order term (which depends on the input
parameters) is exact; and so the rest of the expansion is set to zero.
\ This last step determines the unknown parameters to be substituted
into the zeroth-order term.
\ Note that FAC is not a variational approach as the parameters are
not determined by varying.

A scheme with the same starting point which is variational is the
so-called ``principle of minimal sensitivity" or PMS \cite{Stevenson}.
\ After obtaining a truncated perturbation expansion, one varies it
with respect to the undetermined parameter(s).
\ Note that these parameters were introduced artificially and that the
actual answer should not depend on them; however, any truncated expansion
does depend on them.
\ The PMS philosophy is then to search for the result that is ``least
sensitive" to the parameters -- and hence the variation.
\ In a few select simple examples the PMS procedure has been
proven to yield a convergent series of approximations even when
the underlying perturbation expansion is asymptotic \cite{converge}, but
the general conditions for which it does so remains an open problem.

\subsection{Comparison of Approaches on An Example from Quantum
Mechanics}

Let us test these approaches on a well-known problem from quantum
mechanics, the quartic oscillator: \ $H=-{1 \over 2}d^2/dx^2 +
{g \over 4} x^4$ using the harmonic oscillator $H^{(0)}=-{1 \over 2}
d^2/dx^2 +\frac{1}{2}\omega^2x^2$ as the basis for the perturbation
expansion.
\ By dimensional arguments, the eigenvalues of $H$ can be seen to be
proportional to $g^{1/3}$, and hereafter we scale this factor out.
\ Let us focus our attention on the ground-state energy; the result
is known to be $E_0^{\rm direct}=0.420805 \ldots$.
\ Applying the MWC, FAC and PMS\cite{converge,Neveu} approaches to standard
Rayleigh-Schr\"odinger perturbation theory (calculated to third order
for the energy and second order for wave function) yield the results
seen in Table 1.
\ We also include the Rayleigh-Ritz result and its first Lan\c czos-like
correction (RRL).

Let us make a few observations.
\ PMS and RRL are identical at first order.
\ At second-order the FAC and PMS approaches have no physical solutions;
in PMS one can search for inflection points when no extrema are found ---
at second order this yields $0.42143$.
\ Notice that not only are the variational approaches better than FAC at
first order but also they were improved by going to second or third order,
while FAC got worse.
\ To the orders calculated here, PMS has led to the best results.

\section{Some Fokker-Planck Examples}

In this section we will consider a few examples of the variational
schemes applied to Fokker-Planck problems.
\ First, we will apply the techniques to a toy model.
\ Then we will examine a more difficult problem seen recently in the
physics literature --- the colored-noise problem.

\subsection{A Toy Model}
The toy model we consider is the Fokker-Planck equation corresponding
to the following two coupled, non-linear Langevin equations:
\begin{eqnarray}
{dx_1 \over dt} \ &=& \ -\nu x_1 - g
x_1x_2^2 + \eta_1(t) \nonumber \\
{dx_2 \over dt} \ &=& \ -\nu x_2 + \eta_2(t),
\label{toy-langevin}
\end{eqnarray}
where the noise is Gaussian-correlated with zero mean and the following
correlation:
\begin{equation}
\langle \eta_i(t) \eta_j(t^{\prime}) \rangle \ =\
2D \delta_{ij} \delta(t-t^{\prime}).
\end{equation}
The associated Fokker-Planck operator is:
\begin{equation}
L_{\rm toy} \ = \ D \left({\partial^2 \over \partial x_1^2}+
{\partial^2 \over \partial x_2^2} \right)
+\nu \left(2+x_1{\partial \over \partial x_1}+
x_2{\partial \over \partial x_2}\right)
+g x_2^2\left( 1+x_1{\partial \over \partial x_1}
\right).
\label{toy-fp}
\end{equation}

The first thing to note is that it does not satisfy the potential
conditions.
\ Since they are not satisfied, let us apply each of the variational
methods suggested above to obtain the steady-state values of
$\langle x_1^2 \rangle$ and $\langle x_2^2 \rangle$.
\ For the approaches based on perturbation theory we will
need an operator for which the eigenvalue problem is solvable.
\ We choose as $L_{\rm toy}^{(0)}$ the Fokker-Planck operator for an
Ornstein-Uhlenbeck process (the harmonic oscillator of Fokker-Planck
problems):
\begin{equation}
L_{\rm toy}^{(0)} \ =\ \nu d_1 {\partial^2 \over \partial x_1^2}+
\nu d_2 {\partial^2 \over \partial x_2^2}
+ \nu \left(2+x_1{\partial \over \partial x_1}
+x_2{\partial \over \partial x_2}\right).
\label{OU}
\end{equation}
The eigenstates of $L_{\rm toy}^{(0)}$ are:
\begin{equation}
P_{n_1,n_2}^{(0)}(\underline{x}) \ = \
{H_{n_1}\left(x_1 / \sqrt{2d_1}\right)
H_{n_2}\left(x_2 / \sqrt{2d_2}\right) \over
\left(2^{n_1+n_2+2}\pi^2 n_1! n_2! d_1 d_2 \right)^{1/2}}
\exp\left\{~ -~{x_1^2  \over
2 d_1} ~-~{x_2^2 \over 2 d_2} ~\right\},
\end{equation}
and its eigenvalues are $\lambda_{n1,n2}= -(n_1+n_2)\nu$ where
$n_1,n_2=0,1,2,...$.
\ As we are interested in spacial quantities, we have selected parameters
$d_i$ which affect the spatial distribution $P_0^{(0)}(\underline{x})$
but leave the eigenvalues of $L_{\rm toy}^{(0)}$ unchanged.

The details of the variational calculations we performed on the
toy model can be found in the appendix.
\ The results are shown in Table 2 along with the results from
a simulation of the toy-model Langevin equation (\ref{toy-langevin}).

The simulation algorithm employed a second-order Runge-Kutta method
to evolve the equations. \cite{NR}
\ After a sufficiently long evolution, dependence on the initial
conditions is lost.
\ We simulated the equation for $1~000~000$ realizations
and extracted the averages $\langle x_i^2(t_f) \rangle$ where
$x_i(t_f)$ is the final position of the simulation.
\ The results of a simulation with $D=0.5$, $\nu=1.0$ and
$g=3.0$ are $\langle x_1^2 \rangle = 0.2652280$ and
$\langle x_2^2 \rangle = 0.5003104$.

Note that the values for $d_1$ and $d_2$ found from varying
$\langle \psi_0^{(1)}|\psi_0^{(1)} \rangle$ were substituted
into eq. (\ref{x2-pert}) at first order.
\ The methods based on perturbation theory all agree that
$\langle x_2^2 \rangle =0.5$ and are better than the $\langle
\psi_0^{(0)}|\ell_{\rm toy}^{\dagger}\ell_{\rm toy}|\psi_0^{(0)} \rangle$
variation on this point; on the other hand, the latter variation
produced the best result for $\langle x_1^2 \rangle$. In fact, it is easy
to see, by direct integration of the second equation (\ref{toy-langevin}),
that $\langle x^2_2 \rangle = D/\nu$, which is exactly 0.5 for the values of
the parameters we chose. By beginning with this value at first order, the
methods based on perturbation theory had a distinct advantage over the
$L^{\dagger}L$ approach --- although the values of the variational
parameters $d_1$ and $d_2$ were very similar in the MCW and $L^{\dagger}L$
cases. One reason for the relatively poor result for $\langle x^2_2 \rangle$
when not using the perturbative schemes, may be that the variation in this
case is not carried out directly on the quantity of interest.

\subsection{The Colored-Noise Example}
Now we move on to consider some of the variational approaches for a more
difficult situation --- the colored-noise problem
\cite{review,PathIntI,ProbDist}.
\ Consider the following Langevin equation:
\begin{equation}
\dot x \ =\ -V^{\prime}(x)\ +\ \xi(t),
\label{coloredLangevin}
\end{equation}
which describes an overdamped particle subject to the force
$f(x)=-V^{\prime}(x)$ and an external noise $\xi(t)$.
\ External noise is not intrinsically related to the system's evolution
and is typically ``colored" as opposed to ``white," i.e. it is not
delta-function correlated.
\ As a consequence, many of the techniques and results familiar from
the study of Markov processes are not applicable.
\ For present purposes, we take $\xi(t)$ to be Gaussian-distributed
with zero mean and exponentially correlated:
\begin{equation}
\langle \xi(t) \xi(t^{\prime}) \rangle \ =\ {D \over \tau}
\exp \left\{-|t-t^{\prime}|/\tau \right\}.
\label{noise-cor}
\end{equation}
With this choice, the above one-dimensional non-Markov process
(\ref{coloredLangevin}) can be shown to be equivalent to the
following two-dimensional Markov process \cite{PathIntI}:
\begin{eqnarray}
\dot x \ &=&\ -V^{\prime}(x)\ +\ \xi(t), \nonumber \\
\dot \xi \ &=&\ -{1 \over \tau}\xi \ + \ {1 \over \tau} \eta(t),
\end{eqnarray}
where $\eta(t)$ is Gaussian-distributed with zero mean and
this time delta-function correlated:
\begin{equation}
\langle \eta(t) \eta(t^{\prime}) \rangle \ = \
2D \delta(t-t^{\prime}).
\end{equation}
The corresponding Fokker-Planck equation is:
\begin{equation}
{\partial P(x,\xi ;t) \over \partial t} \ =\
-{\partial \over \partial x} \biggl[\left[-V^{\prime}
(x) + \xi\right]P \biggr] + {1 \over \tau} {\partial
\over \partial \xi} \biggl[ \xi P + {D \over \tau}
{\partial P \over \partial \xi} \biggr],
\label{coloredFPI}
\end{equation}
which upon the transformation to the velocity variable
$\xi=\dot x+ V^{\prime}(x)$ becomes:
\begin{equation}
{\partial P(x,\dot x ;t) \over \partial t} \ =\
\left\{ {D \over \tau^2} {\partial^2 \over \partial \dot x^2}
+ \left[V^{\prime \prime}(x) +{1 \over \tau} \right]
\left[1 + \dot x {\partial \over \partial \dot x} \right]
-\dot x {\partial \over \partial x}
+ \left[{V^{\prime}(x) \over \tau} \right]{\partial \over
\partial \dot x} \right\} P.
\label{coloredFPII}
\end{equation}

Again, it is readily confirmed that the potential conditions
are not satisfied in this case.
\ Nevertheless, there is a solvable case.
\ When the operator $L$ is quadratic (i.e. when the drift vector
$A_i(\underline{x})$ is linear in and the diffusion matrix
$B_{ij}(\underline{x})$ is independent of the $x_i$'s), one can
find a closed expression for the stationary distribution even
if the potential conditions are not met.
\ For the colored-noise problem, this situation occurs if
$V(x)=\omega^2x^2/2$.
\ The associated Fokker-Planck operator is:
\begin{equation}
L_{\rm quad} \ =\
 {D \over \tau^2} {\partial^2 \over \partial \dot x^2}
+ \biggl[\omega^2 +{1 \over \tau} \biggr]
\biggl[1 + \dot x {\partial \over \partial \dot x} \biggr]
-\dot x {\partial \over \partial x}
+ \biggl[{\omega^2 x \over \tau} \biggr]{\partial \over
\partial \dot x} ,
\label{CN-Quad-L}
\end{equation}
and steady-state distribution is:
\begin{equation}
P_0 \ \propto \ \exp \left\{~-~{\omega^2 (1+\omega^2\tau)
\over 2D}x^2 ~-~ {\tau(1+\omega^2 \tau) \over 2D}\dot
x^2 ~\right\} .
\label{CN-Quad-P}
\end{equation}
This result can be useful for considering the behavior near the minima
for a more general potential $V(x)$.

Hereafter, we take $V(x)$ to be a bistable potential given by
$V(x)=-x^2/2+x^4/4$, with the following operator:
\begin{equation}
L_{\rm bistable} \ =\
 {D \over \tau^2} {\partial^2 \over \partial \dot x^2}
+ \left[3x^2-1 +{1 \over \tau} \right]
\left[1 + \dot x {\partial \over \partial \dot x} \right]
-\dot x {\partial \over \partial x}
+ \left[{x^3-x \over \tau} \right]{\partial \over
\partial \dot x} .
\label{CN-B-L}
\end{equation}
For the region immediately surrounding the minima of the bistable
$V(x)$ ($x=\pm 1$), the potential can be described by Taylor
expansion truncated at the quadratic order, suggesting
that
\begin{equation}
\lim_{x \rightarrow 1}
P_0\ \approx \ \exp \left\{~-~ {(1+2\tau) \over
D}(x-1)^2 ~-~ {\tau(1+2 \tau) \over 2D}\dot x^2 ~\right\}
\label{aroundone}
\end{equation}
and similarly around $x=-1$.

As we know of no convenient solvable problem with a bistable
potential, let us take the $L^{\dagger}L$ approach.
\ All we need is some suitable trial steady-state distribution.
We choose a variational stationary distribution of the form:
\begin{equation}
P(x,\dot x) \ \sim \ h(x) ~\exp
\left\{~ -~f(x) ~-~g(x)\dot x^2 ~\right\},
\label{trialP}
\end{equation}
so that the integrations over $\dot x$ are readily performed
analytically.
\ Note that stability requires that $g(x)>0$ for all
$x$.
\ Furthermore, the symmetry suggests that $f(x)$, $g(x)$
and $h(x)$ are even in $x$.
\ Calculating the expectation of $L^{\dagger}L$ with respect
to $P$ yields:
\begin{eqnarray}
{\langle P |L^{\dagger} L |P \rangle \over
\langle P | P \rangle } \ &=&\  \int dx \Biggl\{
{15\left(g^{\prime}(x)\right)^2 \over
64\left(g(x)\right)^{7/2}}
+ {3 \over 4\left(g(x)\right)^{1/2}} \left[
{2D g(x)\over \tau^2} -3x^2+1-{1 \over \tau} \right]^2
\nonumber \\
& &+ {3g^{\prime}(x) \over 8\left(g(x)\right)^{5/2}} \left[
-{h^{\prime}(x) \over h(x)}+
f^{\prime}(x)- {2(x^3-x)g(x) \over \tau} \right] \nonumber \\
& &+  {1 \over 4\left(g(x)\right)^{3/2}} \left[
-{h^{\prime}(x) \over h(x)}+
f^{\prime}(x)- {2(x^3-x)g(x) \over \tau} \right]^2\Biggr\}
\left(h(x)\right)^2\exp \{-2f(x)\} \nonumber \\
& & \times \left[ \int dx {\left(h(x)\right)^2
\over \left(g(x)\right)^{1/2}} \exp
\{-2f(x)\} \right]^{-1}
\label{expect}
\end{eqnarray}
where the $\dot x$ integration has been done.

We use the following simple polynomials for $f(x)$, $g(x)$
and $h(x)$:
\begin{eqnarray}
f(x) &\ =\ & {(1 +2\tau) \over D}\left[
{(x^2-1)^2 \over 4} ~+~ a \tau {(x^2-1)^3 \over 6}
\right]\nonumber \\
g(x) &\ =\ & {\tau (1+2\tau) \over 2D}
\left[1 ~+~ b\tau {(x^2 -1) \over 2} \right] \nonumber \\
h(x) &\ =\ & 1~+~ c\tau{(x^2 -1) \over 2}
\label{fandg},
\end{eqnarray}
where $a$, $b$ and $c$ are variational parameters.
\ This motivation for these choices is the following:
we know what $f(x)$, $g(x)$ and $h(x)$ should be near
$x=\pm 1$, and this variation represents in some sense
the next terms in the Taylor expansions of these functions,
expanding simultaneously around $x=\pm 1$.
\ (It also recovers the appropriate $\tau \rightarrow 0$
limit.)
\ If $D=0.1$ and $\tau=0.6$, we find $\langle P |L^{\dagger} L
|P \rangle=0.32015$ when $a=0,b=0$ and $c=0$, whereas varying $a$, $b$
and $c$ yielded a minimum of $\langle P |L^{\dagger} L |P \rangle=
0.037117$, an order of magnitude smaller, at $a=1.14, b=2.39$ and $c=1.85$.
\ The corresponding marginal stationary distribution
\begin{equation}
P_{mar}(x) \ =\ \int d \dot x P(x,\dot x) \ \sim \
{h(x) \over [g(x)]^{1/2}}\exp \left\{ -f(x) - {\rm ln}\left[g(x)\right]/2 +
{\rm ln}\left[h(x)\right] \right\},
\label{margP}
\end{equation}
is shown in Figure 1.
\ Note that while there is certainly room for improvement, it does
capture some of the features such as the shift in the maxima away from
$x=\pm 1$, due to existence of $x$-dependent prefactors.

These results can be compared to those obtained by a systematic small-$D$
expansion using path-integral techniques.
\ The path-integral approach is clearly superior for small $D$, but we
expect it to become less reliable as $D$ increases.
\ In contrast, the variational approach can, in principle, be applied
for any $D$.

\section{Conclusions}

In this paper we have discussed a number of variational procedures which
may be applied to the Fokker-Planck equation.
\ They broadly fall into two classes: \ those in which a Hermitian
operator was constructed from the generically non-Hermitian Fokker-Planck
operator and those which were constructed with reference to an
approximation to the full problem, which was exactly soluble.
\ Both approaches have advantages and disadvantages.
\ For example, in the former approach, where we used the Hermitian operator
$L^{\dag}L$, twice as many operations have to be carried out --- the result
of first operating with $L$ and then with $L^{\dag}$.
\ On the other hand, this may be the only method available; in the
colored noise example, for instance, we were unable to construct a
Fokker-Planck operator which was both solvable and which contained enough
of the essential physics to make it a useful approximation.
\ The variants in the second approach were tested on a simple
quantum-mechanical example as well on a toy Fokker-Planck problem.
\ One of these schemes (MCW) shares with $L^{\dagger}L$ the advantage of
having a lower bound built in.
\ From the few examples we have investigated, it is clear that
variational procedures are capable of giving good quantitative, as well
as qualitative, results for Fokker-Planck equations.
\ However more work is required to extend the range of the problems
studied, as well as the order to which the calculations are carried out,
before definitive statements about the relative usefulness of the various
possible approaches can be made.

\acknowledgements

We wish to thank N. Goldenfeld and M. Moore for useful discussions and the
Isaac Newton Institute for Mathematical Sciences, where this work was begun,
for its hospitality. TB acknowledges the support of the Engineering and
Physical Sciences Research Council under grant GR/H40150.

\appendix
\section*{}

In this appendix, some of the calculational details for the toy model
are provided.
\ The perturbative approaches split the operator $L_{\rm toy}$,
eq. (\ref{toy-fp}), as follows $L_{\rm toy}=L_{\rm toy}^{(0)}
+L_{\rm toy}^{(1)}$,
where $L_{\rm toy}^{(0)}$, eq. (\ref{OU}), is the Fokker-Planck operator
for the Ornstein-Uhlenbeck process.
\ The associated $L_{\rm toy}^{(1)}$ is:
\begin{equation}
L_{\rm toy}^{(1)} \ =\  \sum_{i=1,2}(D-\nu d_i)
{\partial^2 \over \partial x_i^2}
+g x_2^2 \left( 1+x_1{\partial \over \partial x_1}
\right).
\end{equation}
It is convenient to work with the transformed operator
$\left(P_0^{(0)}(\underline{x})\right)^{-1/2}L\left(P_0^{(0)}
(\underline{x})\right)^{1/2}$; expressed in terms of annihilation
and creation operators, it is:
\begin{equation}
\ell_{\rm toy}^{(0)} \ = \ -\nu ( a_1^{\dagger} a_1 + a_2^{\dagger} a_2 )
\end{equation}
\begin{equation}
\ell_{\rm toy}^{(1)} \ = \ \sum_{i=1,2} \left({D \over d_i}-\nu\right)
a_i^{\dagger} a_i^{\dagger} -g  d_2 (a_2 + a_2^{\dagger})^2 (a_1^{\dagger}
a_1 + a_1^{\dagger} a_1^{\dagger}),
\end{equation}
where $x_i=d_i^{1/2}(a_i + a_i^{\dagger})$ and $\partial/\partial
x_i=(a_i - a_i^{\dagger})/2d_i^{1/2}$.

{}From here one can readily calculate the following:
\begin{equation}
\langle \psi_0^{(0)}|\ell_{\rm toy}^{\dagger}\ell_{\rm toy}
|\psi_0^{(0)} \rangle \ = \
2\left[{D \over d_1} - \nu - g d_2 \right]^2
+2\left[{D \over d_2} - \nu \right]^2
+4 g^2 d_2^2 ,
\end{equation}
and similarly
\begin{equation}
\langle \psi_0^{(1)}|\psi_0^{(1)} \rangle \ = \
{1 \over 2  \nu^2}
\left[{D \over  d_1} - \nu - g d_2 \right]^2
+{1 \over 2  \nu^2}
\left[{D \over d_2} - \nu \right]^2
+{g^2 d_2^2 \over  4\nu^2 }.
\end{equation}

Minimization of $\langle \psi_0^{(0)}|\ell_{\rm toy}^{\dagger}
\ell_{\rm toy}|\psi_0^{(0)} \rangle$ yields:
\begin{equation}
d_1 \ =\ {D \over \nu +g d_2} \ \ \
{\rm and} \ \ \
d_2^4 + {D\nu \over 2g^2} d_2 - {D^2 \over 2g^2} = 0;
\end{equation}
while minimization of $\langle \psi_0^{(1)}|\psi_0^{(1)} \rangle$
results in:
\begin{equation}
d_1 \ =\ {D \over \nu +g d_2} \ \ \
{\rm and} \ \ \
d_2^4 + {2D\nu \over g^2} d_2 - {2D^2 \over g^2} = 0.
\end{equation}

To include the FAC and PMS approaches, we calculate $\langle
x_i^2 \rangle$:
\begin{equation}
\langle x_i^2 \rangle \ =\  \int dx_1 dx_2 ~ x_i^2 ~
\left(P_0^{(0)}({\underline x}) + \epsilon P_0^{(1)}({\underline x})
+ \epsilon^2 P_0^{(2)}({\underline x}) + \ldots \right),
\end{equation}
for $i=1,2$.
\ To order $\epsilon^2$ we find:
\begin{eqnarray}
\langle x_1^2 \rangle \ =\ && d_1 + \epsilon\left[{d_1 \over \nu }
\left({D \over d_1} -\nu -gd_2\right) \right] \nonumber \\
&&+ \epsilon^2 \left[{d_1 \over \nu } \left({2g^2d_2^2 \over \nu}
+ 2gd_2 -{gd_2D \over \nu d_1}-{gD \over \nu} \right) \right]
\nonumber \\
\langle x_2^2 \rangle \ =\ && d_2 + \epsilon\left[{d_2 \over \nu }
\left({D \over d_2} -\nu \right) \right] +\epsilon^2 \left[ 0 \right].
\label{x2-pert}
\end{eqnarray}
FAC gives $\langle x_1^2 \rangle = D\nu/(\nu^2 + gD)$ and
$\langle x_2^2 \rangle = D/\nu$ at $O(\epsilon)$, and has no
physical solution at $O(\epsilon^2)$.
\ PMS, on the other hand, has no physical solution at $O(\epsilon)$
and variation of $\langle x_1^2 \rangle$ at $O(\epsilon^2)$ with
respect to $d_1$ and $d_2$ yields:
\begin{eqnarray}
&&{2g^2 d_2^2 \over \nu^2} + {g d_2 \over \nu} - {g D \over \nu^2}
= 0 \nonumber \\
&&{4g^2 d_1 d_2 \over \nu^2} + {g d_1 \over \nu} - {g D \over \nu^2}
= 0.
\end{eqnarray}

\newpage

\vbox{
\centerline{{\bf Table 1}}
\begin{displaymath}
\begin{tabular}{|c|c|c|c|c|}
\hline
\hline
\ \ Order\ \  & MWC & FAC & PMS & RRL \\
\tableline
1 & \ \ 0.42957 \  \ & \ \ 0.45428\ \  & \ \ 0.42927 \ \
& \ \ 0.42927 \ \ \\
2 & 0.42242 & --- & (0.42143) & 0.42236 \\
3 & --- & 0.48854 & 0.42098 & --- \\
\hline
\hline
\end{tabular}
\end{displaymath}
\centerline{\small
The results of four variational approaches applied to $E_0$ of
the quartic oscillator.}
}

\newpage

\vbox{
\centerline{{\bf Table 2}}
\begin{displaymath}
\begin{tabular}{|c|c|c|}
\hline \hline
\ Method \ & $\langle x_1^2\rangle$ & $\langle x_2^2\rangle$ \\
\tableline
\ Simulation \ & \ \ 0.265228 \ \ & \ \ 0.500310 \ \ \\
$L^{\dagger}L$ & 0.271875 & 0.279693 \\
MCW & 0.241854 & 0.5 \\
FAC & 0.2 & 0.5 \\
PMS & 0.174307 & 0.5 \\
\hline \hline
\end{tabular}
\end{displaymath}
\centerline{\small Variational results for the toy model with $D=0.5$,
$v=1.0$ and $g=3.0$. }
}

\begin{figure}
\end{figure}

\begin{description}

\item{Fig. 1} The value of $ln[P_st(x)]$ plotted for $D=0.1$ and
$\tau=0.6$.  The dots are from a numerical simulation of the
Langevin equation and the solid curve is the result of the variational
calculation.

\end{description}

\end{document}